\documentclass[sigplan,screen]{acmart}

\AtBeginDocument{%
  }

\copyrightyear{2025} 
\acmYear{2025} 
\setcopyright{acmlicensed}
\acmConference[ASPLOS '25]{Proceedings of the 30th ACM International Conference on Architectural Support for Programming Languages and Operating Systems, Volume 2}{March 30-April 3, 2025}{Rotterdam, Netherlands} 
\acmBooktitle{Proceedings of the 30th ACM International Conference on Architectural Support for Programming Languages and Operating Systems, Volume 2 (ASPLOS '25), March 30-April 3, 2025, Rotterdam, Netherlands} 
\acmISBN{979-8-4007-1079-7/25/03}
\acmDOI{10.1145/3676641.3715986} 

\usepackage[]{hyperref}
\renewcommand{\sectionautorefname}{\S\kern-0.2em}
\renewcommand{\subsectionautorefname}{\S\kern-0.2em}

\usepackage{multirow}
\usepackage{algorithm}
\usepackage{algorithmic}
\usepackage{setspace}
\usepackage{soul}
\usepackage{xcolor}
\usepackage{tabularx}
\usepackage{url}
\usepackage{balance}
\sethlcolor{yellow}

\newcommand{\liao}{\textcolor{black}}

\begin{document}

\title[CoServe: Efficient Collaboration-of-Experts (CoE) Model Inference]{CoServe: Efficient Collaboration-of-Experts (CoE) Model Inference with Limited Memory}

\author{Jiashun Suo}
\orcid{0000-0002-5360-353X}
\email{suojiashun@buaa.edu.cn}
\affiliation{%
  \institution{State Key Laboratory of CCSE and School of Computer Science and Engineering, Beihang University}
  \city{Beijing}
  \country{China}
}

\author{Xiaojian Liao}
\authornote{Xiaojian Liao and Limin Xiao are corresponding authors.}
\orcid{0000-0002-7924-9268}
\email{liaoxj@buaa.edu.cn}
\affiliation{%
  \institution{State Key Laboratory of CCSE and School of Computer Science and Engineering, Beihang University}
  \city{Beijing}
  \country{China}
}

\author{Limin Xiao}
\authornotemark[1]
\orcid{0000-0001-9438-9181}
\email{xiaolm@buaa.edu.cn}
\affiliation{%
  \institution{State Key Laboratory of CCSE and School of Computer Science and Engineering, Beihang University}
  \city{Beijing}
  \country{China}
}

\author{Li Ruan}
\orcid{0000-0002-2386-961X}
\email{ruanli@buaa.edu.cn}
\affiliation{%
  \institution{State Key Laboratory of CCSE and School of Computer Science and Engineering, Beihang University}
  \city{Beijing}
  \country{China}
}

\author{Jinquan Wang}
\orcid{0000-0001-6690-8386}
\email{derekjqwang@buaa.edu.cn}
\affiliation{%
  \institution{State Key Laboratory of CCSE and School of Computer Science and Engineering, Beihang University}
  \city{Beijing}
  \country{China}
}

\author{Xiao Su}
\orcid{0000-0001-5365-2537}
\email{xiaosu@buaa.edu.cn}
\affiliation{%
  \institution{State Key Laboratory of CCSE and School of Computer Science and Engineering, Beihang University}
  \city{Beijing}
  \country{China}
}

\author{Zhisheng Huo}
\orcid{0000-0002-5366-0892}
\email{huozhisheng1122@126.com}
\affiliation{%
  \institution{State Key Laboratory of CCSE and School of Computer Science and Engineering, Beihang University}
  \city{Beijing}
  \country{China}
}

\renewcommand{\shortauthors}{Jiashun Suo, Xiaojian Liao, Limin Xiao et al.}

\begin{abstract}
Large language models like GPT-4 are resource-intensive, but recent advancements suggest that smaller, specialized experts can outperform the monolithic models on specific tasks.
The Collaboration-of-Experts (CoE) approach integrates multiple expert models, improving the accuracy of generated results and offering great potential for precision-critical applications, such as automatic circuit board quality inspection.
However, deploying CoE serving systems presents challenges to memory capacity due to the large number of experts required, which can lead to significant performance overhead from frequent expert switching across different memory and storage tiers.

We propose CoServe, an efficient CoE model serving system on heterogeneous CPU and GPU with limited memory. 
CoServe reduces unnecessary expert switching by leveraging expert dependency, a key property of CoE inference.
CoServe introduces a dependency-aware request scheduler and dependency-aware expert management for efficient inference.
It also introduces an offline profiler to automatically find optimal resource allocation on various processors and devices. 
In real-world intelligent manufacturing workloads, CoServe achieves 4.5$\times$ to 12$\times$ higher throughput compared to state-of-the-art systems.

\end{abstract}

\ccsdesc[500]{Computer systems organization~Architectures}
\ccsdesc[500]{Computing methodologies~Artificial intelligence}
\ccsdesc[300]{Computer systems organization~Real-time systems}

\keywords{Collaboration-of-Experts (CoE); ML Inference; Edge Computing}

\maketitle 

\section{Introduction}
Large language models (LLMs), such as GPT-4~\cite{achiam2023gpt}, have revolutionized modern AI applications.
Traditionally, LLMs are monolithic, each encompassing billions or even trillions of parameters, demanding significant costs for both training and fine-tuning.
However, recent advances in machine learning research have shown that smaller experts~\footnote{For brevity, this paper uses the term `experts' to refer to expert models specialized in a particular domain.}, such as Code Llama-Python 7B~\cite{roziere2023code} and Flan-T5-XL 3B~\cite{chung2024scaling}, can outperform general monolithic LLMs on specialized tasks, delivering higher performance and improved accuracy.

The collaboration of multiple experts unlocks vast potential for scenarios demanding high precision in generated outcomes~\cite{jiang2023llm, CoEApp2, lu2023routing, alikhani2023dynafuse, nakano2011multi}.
For instance, in intelligent manufacturing, automated circuit board inspection requires 99.9\% detection accuracy, a level unattainable by a single model.
Fortunately, the \textbf{Collaboration-of-Experts (CoE)} approach, where each expert is trained for distinct components, meets these stringent requirements.

CoE requires substantial memory, posing significant challenges for deployment, particularly on memory-constrained edge devices.
For example, the previously mentioned circuit board inspection application involves over 300 experts (13B parameters, 60GB memory) and must run on edge devices for low latency and privacy.
A typical edge device, such as one with an RTX 3080Ti GPU (12GB memory), cannot fit all experts in GPU memory.
Thus, experts are offloaded to CPU memory or SSD, dynamically loaded into GPU memory during inference as needed.

\begin{figure}[t]
\centering
\includegraphics[width=1\linewidth]{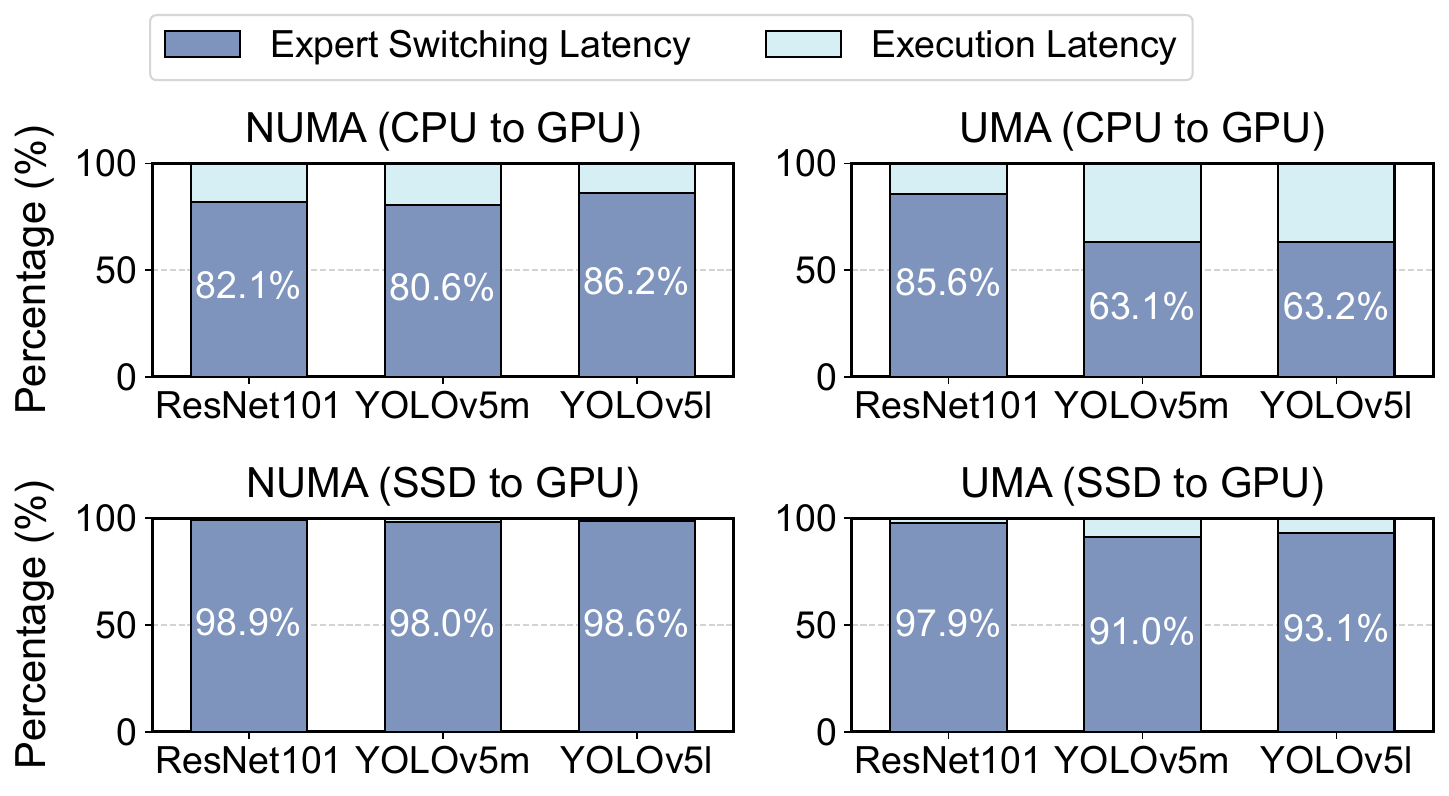}
\caption{Proportion of expert switching latency and execution latency on devices with non-uniform memory architecture (NUMA) and uniform memory architecture (UMA).
The SSD in the NUMA system is a MICRON MTFDDAK480TDS, with a read bandwidth of 530 MB/s.
In the UMA system, the SSD is an APPLE SSD AP0512Z, with a read bandwidth of approximately 3000 MB/s.}
\label{fig1_latency_breakdown}
\end{figure}

However, frequent expert switching incurs significant overhead, severely degrading inference performance.
\autoref{fig1_latency_breakdown} illustrates the proportion of expert switching latency relative to inference latency across various expert types, memory architectures, and I/O paths.
Switching experts from SSD to GPU accounts for over 90\% of inference latency on both NUMA and UMA devices.
Even in unified memory architectures, transferring experts from CPU to GPU consumes over 60\% of inference latency, possibly due to data reorganization by AI frameworks (e.g., PyTorch).
Thus, minimizing expert switching is crucial for efficient CoE serving.

{\color{black}
We initially explored expert management methods from MoE (Mixture of Experts) but found that directly applying these methods led to inefficient expert switching.
The root cause is that expert management in existing MoE models relies on historical statistics (e.g., LRU), with low prediction accuracy, making it ineffective at precisely evicting experts least likely to be used.}


Through further in-depth analysis (\autoref{sec_motivation}), we identify that expert dependency, a key characteristic of CoE systems, is overlooked in existing CoE model serving frameworks, leading to unnecessary expert switching during inference.
First, multiple requests may rely on the same expert, yet their positions in the queue could be spaced far apart.
The state-of-the-art system, Samba-CoE~\cite{prabhakar2024sambanova}, uses a first-come, first-served (FCFS) scheduling strategy.
After the first request requiring expert 1 is completed, the second request, which does not need expert 1, may evict it to SSD.
If the third request depends on expert 1, the system must switch it back into GPU memory.
However, this expert switching could be avoided by reordering the second and third requests.

Second, dependencies can also exist between experts, where subsequent experts in an inference pipeline rely on the output of earlier ones.
Samba-CoE uses the LRU (Least Recently Used)~\cite{lee2001lrfu} strategy to evict experts from GPU memory, but this is inefficient as it considers only historical usage. 
In contrast, CoE models can leverage pre-assessed usage probabilities and dependency relationships among experts, enabling more accurate and efficient expert management.

Moreover, determining the optimal memory allocation for experts in a CoE system is a complex task, especially on edge devices with varying architectures, and different computation and memory capabilities.
Allocating more memory for storing experts reduces the frequency of expert switching but may limit the number of requests that can be processed concurrently (i.e., batch size)~\cite{kwon2023efficient, agrawal2024taming}, leading to underutilization of computational resources.
This tradeoff becomes even more intricate for CoE systems that leverage both GPU and CPU, given their significant differences in computational power and memory capacity.

We introduce CoServe (\autoref{subsec_structure}), an efficient CoE model serving system on heterogeneous CPU and GPU with limited memory.
{\color{black}Unlike existing expert management methods that rely on historical statistical information}, the key idea of CoServe is to exploit expert dependency during inference scheduling and expert management, thereby reducing unnecessary expert switching.
Inspired by our earlier analysis, CoServe introduces three techniques: dependency-aware request scheduling, dependency-aware expert management, and an optional offline profiler.

The dependency-aware request scheduling (\autoref{subsec_request_scheduling}) groups requests that rely on the same expert closer together within a defined window, minimizing the frequency of expert switching. 
It also dynamically distributes requests across different CPU and GPU queues, balancing the workload among queues to achieve lower task-level latency.

The dependency-aware expert management (\autoref{subsec_expert_management}) prioritizes evicting independent experts during expert switching, maximizing the use of limited CPU and GPU memory. 
Additionally, it utilizes pre-assessed usage probabilities for each expert rather than relying on historical statistics, thereby reducing the incidence of unnecessary expert switching.

To automatically adapt to different devices and identify the optimal configuration, the offline profiler is conducted once for each device before system initialization. 
By analyzing expert performance on the device through microbenchmarks, it determines suitable memory allocations (\autoref{subsec_memory_allocation}) during initialization and provides a performance matrix (\autoref{subsec_perf_profiler}) for each expert to enhance request scheduling.

We implement CoServe using PyTorch and deploy it on both NUMA and UMA devices.
We evaluate CoServe in real-world intelligent manufacturing workloads (\autoref{sec_evaluation}), which use multiple experts to detect different types of circuit board defects collaboratively.
Compared to the state-of-the-art system (Samba-CoE), CoServe delivers a 4.5$\times$ to 12$\times$ improvement in terms of throughput.
Ablation studies show that each of CoServe's techniques significantly increases efficiency.

{\color{black}To the best of our knowledge, CoServe is the first system to optimize the inference efficiency of CoE models from a system-level perspective by effectively leveraging scheduling and memory management.}
Our contributions can be summarized as follows: 
\begin{itemize}
    \item We conduct a study on designing an efficient CoE model serving system and identify expert dependency, which is a unique property of CoE, as a crucial factor in reducing expert switching frequency.
    \item We design and implement CoServe, an efficient CoE model serving system on heterogeneous CPU and GPU, empowered by a set of techniques.
    \item We evaluate CoServe in real-world production workloads, which demonstrates significant performance improvement compared to the state-of-the-art system.
\end{itemize}

\section{Background}\label{sec_background}

\begin{figure}[t]
\centering
\includegraphics[width=1\linewidth]{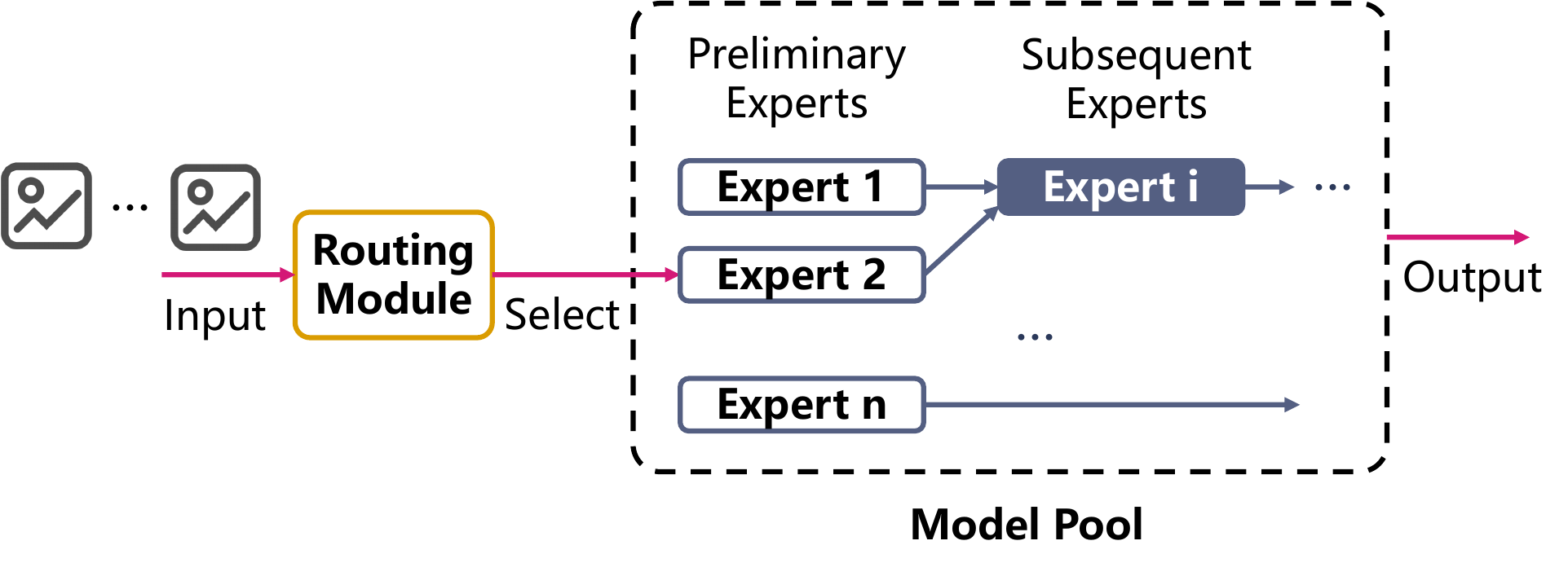}
\caption{Diagram of the Collaboration of Experts (CoE) model inference process.}
\label{fig2_detection_process}
\end{figure}

\subsection{CoE Model and Its Use Cases} \label{subsec_coe_use_case}

CoE (Collaboration or Composition of Experts) integrates multiple expert models to accomplish a task.
As shown in ~\autoref{fig2_detection_process}, it consists of a routing module and several expert models.
{\color{black}The routing module can be configured with user-defined rules or trained independently.}
Upon receiving an input (e.g., an image), it selects a preliminary expert for the first inference.
The output then either guides the next expert selection or directly produces the final result.

Compared to monolithic models, the modular structure of CoE offers several advantages.
First, CoE can achieve higher accuracy.
While a single deep model for circuit board defect detection yields under 92\% accuracy, the collaboration of multiple models increases accuracy to 99.9\%.

Second, CoE simplifies training, as experts can be independently trained and fine-tuned, with the routing module configured manually.
In contrast, MoE (Mixture of Experts)~\cite{jacobs1991adaptive} requires joint training and fine-tuning of both experts and routers, making optimization challenging due to its large parameter count~\cite{rajbhandari2022deepspeed, he2022fastermoe, zhu2024llama}.

{\color{black}
Third, CoE enables more efficient expert management by leveraging an independent routing module to accurately capture expert dependencies and compute expert usage probabilities.
In contrast, MoE routing determines its output only at runtime, forcing it to rely solely on historical data for estimating expert usage probabilities and dependencies, which results in inherent inaccuracies.}

Fourth, CoE offers greater flexibility and scalability, allowing the integration of diverse expert types to seamlessly achieve multi-model capabilities.

A key application of CoE is circuit board defect detection in intelligent manufacturing.
The primary method, Automated Optical Inspection~\cite{liao2018guidelines}, achieves only 80–90\% accuracy, while fully automated detection requires 99.9\%.
A single model is impractical, as its accuracy is limited to 92\% due to component diversity.
Fortunately, CoE improves accuracy to 99.98\%, enabling full automation.
In particular, each component has a dedicated classification expert to detect defects like damage.
If no issues are found, an object detection expert verifies alignment and soldering direction.
Multiple classification experts may share the same object detection expert (e.g., Expert i in ~\autoref{fig2_detection_process}), illustrating shared expert usage.

Another example is Qihoo 360, which uses CoE to combine state-of-the-art models from different domains~\cite{huang2024ccoe}, such as code, math, and law.
By analyzing user requests, the system selects the most appropriate model to handle the corresponding tasks.
The results show that it can easily and efficiently boost performance by nearly 10\%-20\% compared to a single large model across different domains, while using significantly fewer resources for both training and inference.

\subsection{CoE Expert Offloading}

GPU devices with limited memory (e.g., RTX3080Ti with 12GB GPU memory) may be unable to accommodate all CoE experts in GPU memory.
For instance, circuit board defect detection tasks may require the collaboration of over 300 experts and demand memory capacities exceeding 60~GB.

To run CoE models on hardware with limited GPU memory, experts can be offloaded to CPU memory or disks, loading them into GPU memory as needed. 
This enables inference of CoE models of various sizes, provided there is sufficient CPU memory and disk space.
For example, Samba-CoE~\cite{prabhakar2024sambanova} stores frequently used experts in high-bandwidth memory (HBM) and offloads others to DDR. 
When an expert is missing from HBM, the system uses an LRU strategy to swap it from DDR, triggering an expert switch.

\section{Motivation} \label{sec_motivation}

\begin{figure}[t]
\centering
\includegraphics[width=1\linewidth]{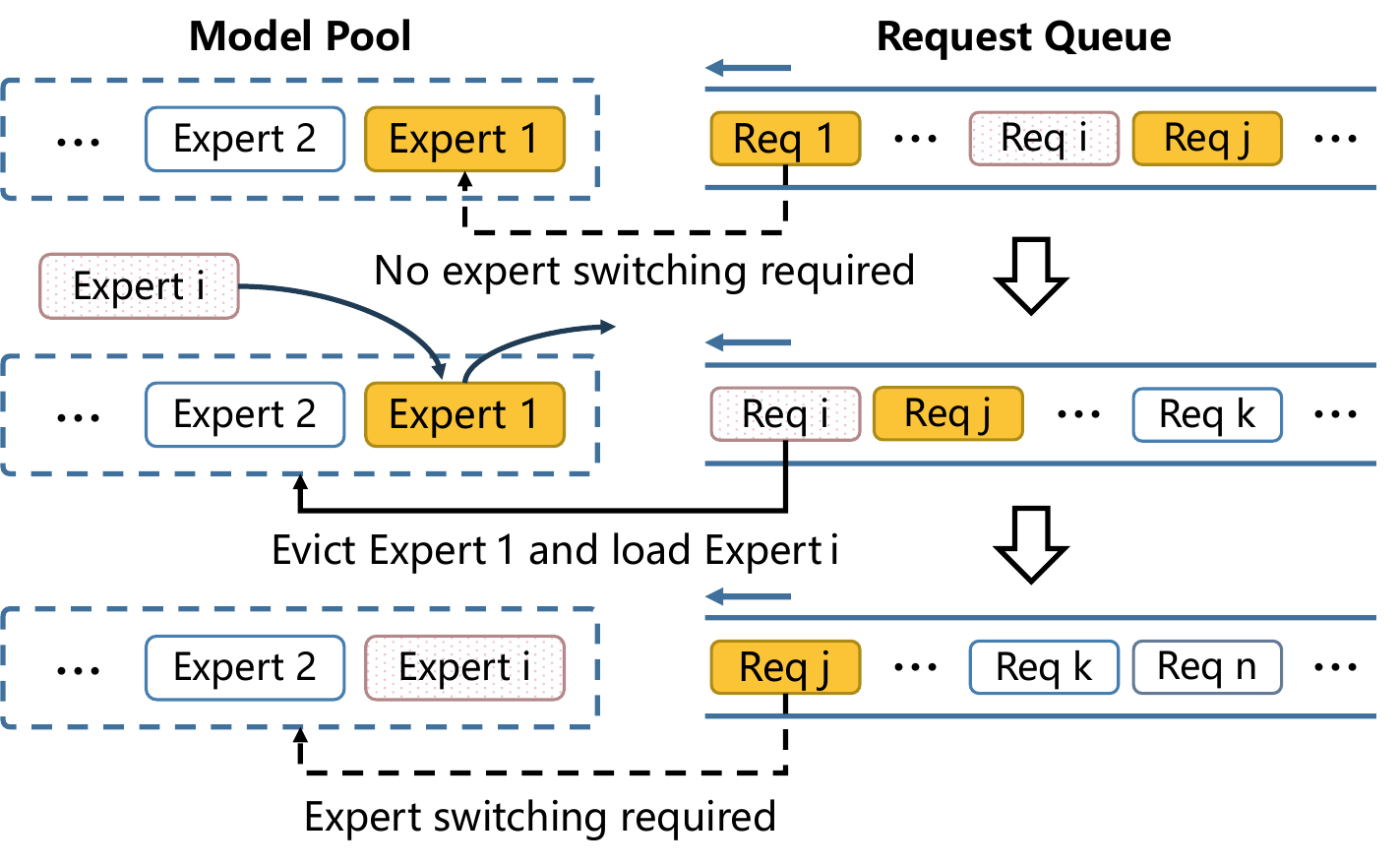}
\caption{Example of expert switching caused by the first-come, first-served approach.}
\label{fig3_cold_start_required}
\end{figure}

{\color{black} Many applications require deploying models at the edge due to privacy and real-time processing needs. For example, in intelligent circuit board inspection, data must not be transmitted outside the factory. 
Additionally, if cloud-based inspection is used, data transmission latency (e.g., >50ms) may exceed edge inference latency (<40ms) for a single image, reducing overall detection efficiency.
}

Edge devices typically have limited memory, leading to frequent expert switching when the number of experts is large.
For instance, in intelligent manufacturing facilities, setups such as an RTX3080Ti with 12 GB of GPU memory or a Jetson Xavier NX with 16 GB often require storing experts on disk and dynamically loading them for inference, which causes expert switching.

However, the expert switching approach significantly degrades inference performance.
\autoref{fig1_latency_breakdown} shows that expert switching latency (from SSD to GPU) accounts for over 90\% of the total inference latency on GPUs for both NUMA (RTX 3080 Ti) and UMA (Apple M2) devices.
Therefore, reducing the frequency of expert switching is crucial for improving inference efficiency, which is the primary focus of this study.

In this section, we thoroughly analyze the inefficiencies of the current CoE model inference system, Samba-CoE, and identify both the challenges and opportunities in designing a more efficient CoE inference system, with a focus on request scheduling (\autoref{subsec_request_scheduling}), expert management (\autoref{subsec_expert_management}), and memory management (\autoref{subsec_memory_allocation}).

\subsection{Request Scheduling}
Samba-CoE processes inference requests on a first-come, first-served basis.
However, more effective request scheduling can avoid some of the resulting expert switching.
For example, as shown in \autoref{fig3_cold_start_required}, both Request 1 and Request j require Expert 1 for inference.
When processing Request 1, Expert 1 is already loaded, eliminating the need for expert switching. 
However, when Request i is processed next, Expert 1 is evicted to free memory for loading Expert i.
Subsequently, when processing Request j, Expert 1 is no longer loaded, requiring expert switching.
If Request j is scheduled to be processed immediately after Request 1, this expert switching can be avoided.

\begin{figure}[t]
\centering
\includegraphics[width=1\linewidth]{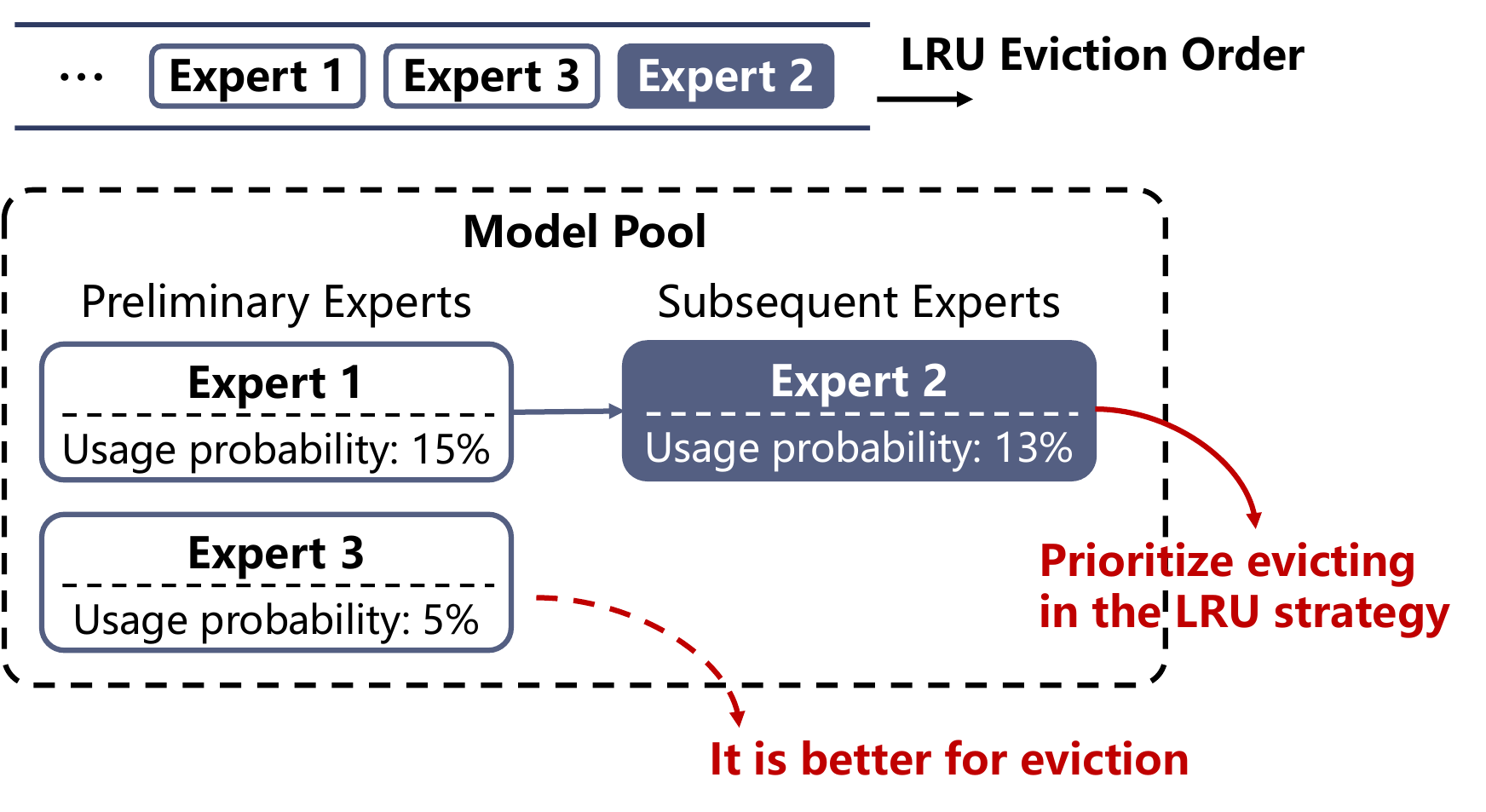}
\caption{Example of expert eviction using the Least Recently Used (LRU) strategy.}
\label{fig4_lru_unload}
\end{figure}

\subsection{Expert Management}

Samba-CoE uses the Least Recently Used (LRU) strategy to manage experts.
LRU relies on past usage data to predict each expert's future demand.
However, these predictions can sometimes be inaccurate.
Specifically, in one application scenario, an expert’s usage probability tends to remain relatively stable due to the consistent data distribution~\cite{eliseev2023fast}, leading to more accurate estimates of future demand.
{\color{black}Because the CoE model’s routing can be either user-defined or trained independently, we can leverage this routing mechanism to calculate each expert’s usage probability based on actual data.}
As shown in \autoref{fig4_lru_unload}, when using the LRU strategy for expert eviction, Expert 2 is evicted first, even though its usage probability is higher than that of Expert 3.
In this case, evicting Expert 3 would be more appropriate.
Therefore, when selecting a candidate expert to evict from GPU memory, relying on the pre-assessed usage probability can yield higher efficiency than the LRU approach.


\begin{figure}[t]
\centering
\includegraphics[width=1\linewidth]{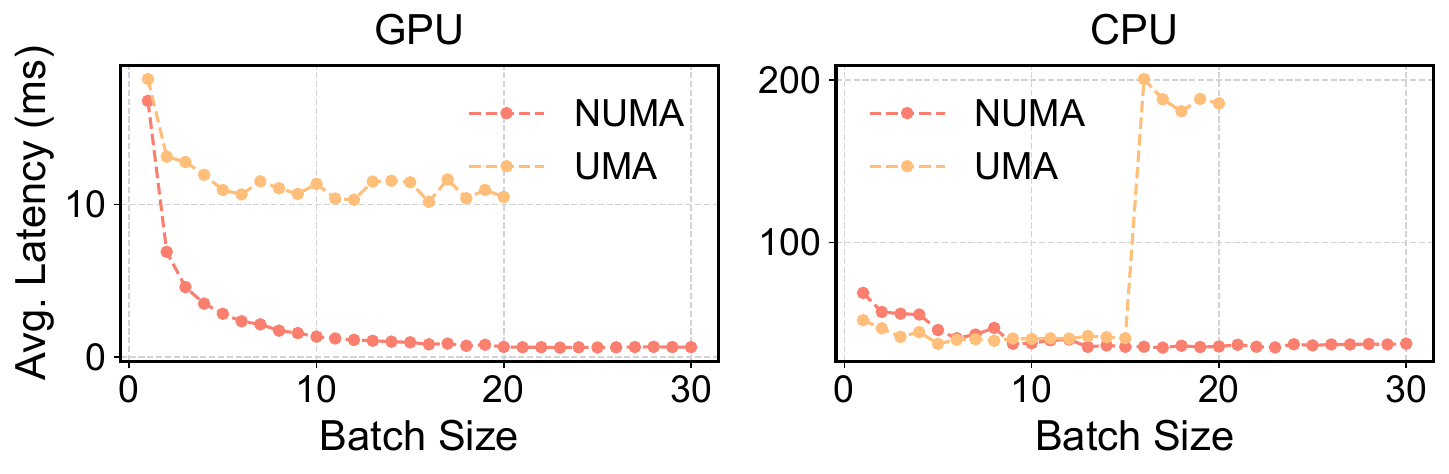}
\caption{Trends in average inference latency with increasing batch size on NUMA and UMA devices.}
\label{fig3_batch_latency_example}
\end{figure}

\subsection{Memory Management} \label{subsec_motiv_memory}

\begin{figure}[t]
\centering
\includegraphics[width=1\linewidth]{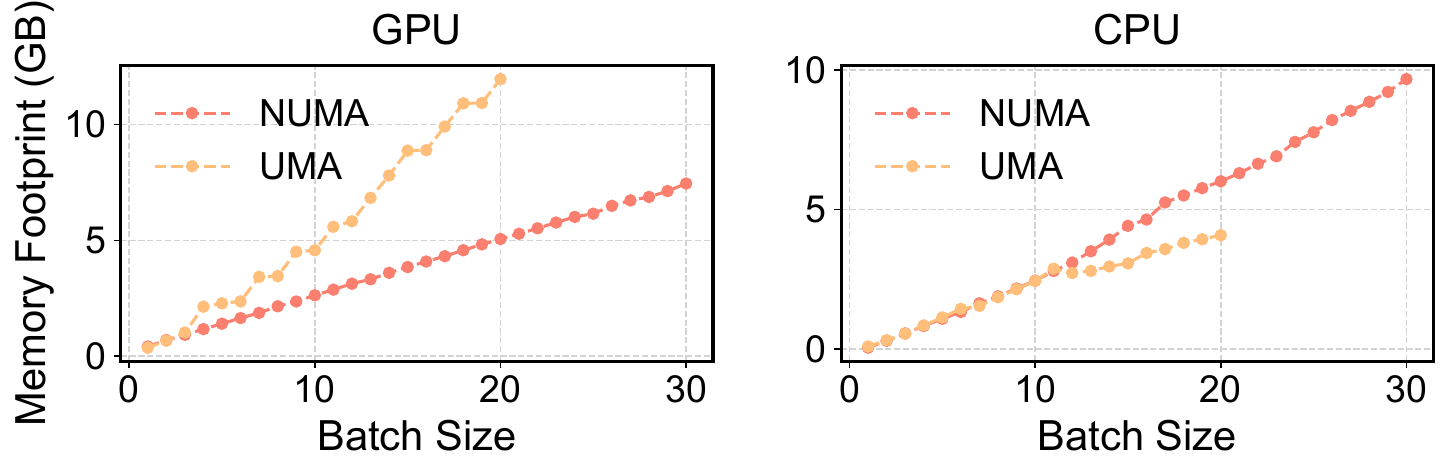}
\caption{Trends in memory footprint with increasing batch size on NUMA and UMA devices.}
\label{fig3_batch_memory_example}
\end{figure}

Memory for experts is divided into two parts: storing expert parameters and intermediate inference results.
Batching multiple requests improves inference performance, as shown in \autoref{fig3_batch_latency_example}, where larger batch sizes reduce average latency (execution latency divided by batch size).
However, beyond a point, benefits diminish.
Increasing batch size also raises memory usage for intermediate results (\autoref{fig3_batch_memory_example}).
For example, increasing ResNet101’s batch size by one consumes as much memory as loading 1.5 experts on a NUMA GPU.
This reduces the number of experts that can be stored in GPU memory.
As a result, the frequency of expert switching may increase.
Thus, balancing memory usage between expert loading and intermediate results remains a key challenge.

Devices often have both CPU and GPU, each exhibiting different performance characteristics and memory footprint depending on the batch size.
For example, on a UMA device, GPU inference achieves the lowest average latency at a batch size of 6, while the CPU performs optimally at a batch size of 5 (\autoref{fig3_batch_latency_example}).
The memory footprint also varies between the CPU and GPU even with the same batch size (\autoref{fig3_batch_memory_example}), due to different data organization methods used by AI frameworks on each.
Similar patterns are consistently observed on NUMA devices as well.
Additionally, the diverse architectures, computational power, and memory capabilities across different devices lead to varying optimal batch sizes and configurations for CoE inference.
This diversity complicates the aforementioned memory tradeoff between expert loading and intermediate results.


\section{CoServe} \label{sec_coserve}

We introduce CoServe, an efficient Collaboration-of-Experts (CoE) model serving system specifically designed for devices with limited memory.
The key idea of CoServe is to reduce the frequency of expert switching by leveraging expert dependencies.
In this section, we first provide a comprehensive overview of CoServe, followed by a detailed explanation of the techniques proposed.

\begin{figure*}[h]
\centering
\includegraphics[width=1\linewidth]{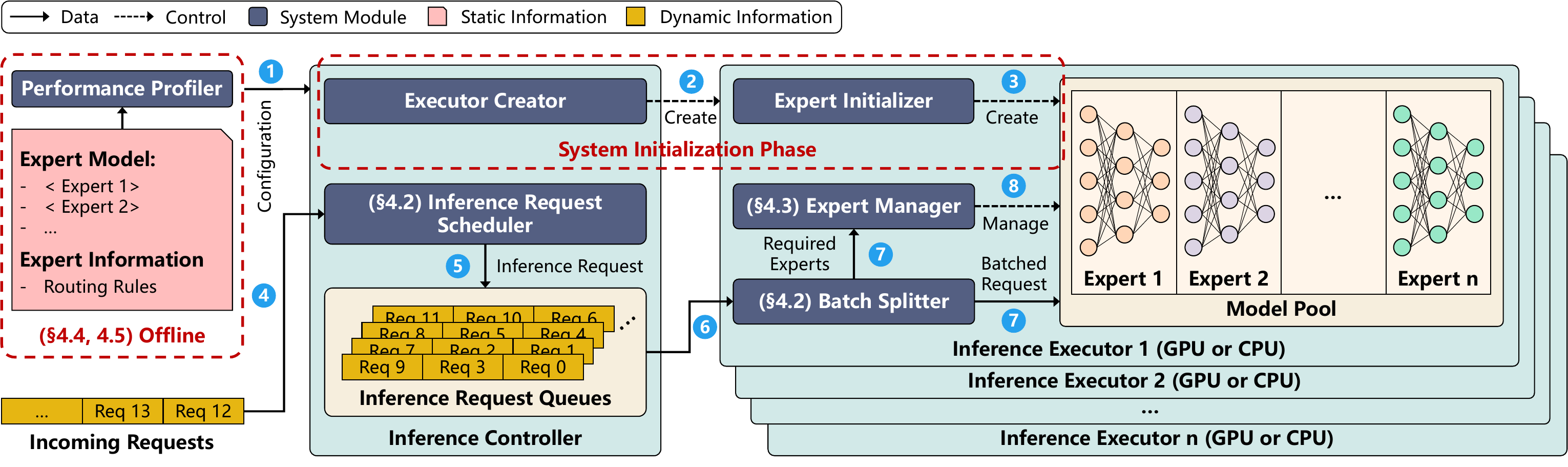}
\caption{CoServe architecture overview.}
\label{fig4_system_structure}
\end{figure*}

\subsection{CoServe Overview} \label{subsec_structure}
\autoref{fig4_system_structure} illustrates the overall architecture of CoServe, which operates in three phases: offline, system initialization, and online.
During the online phase, dependency-aware request scheduling (\autoref{subsec_request_scheduling}) and dependency-aware expert management (\autoref{subsec_expert_management}) are proposed to minimize the frequency of expert switching.
In the offline phase, the optimal memory allocation (\autoref{subsec_memory_allocation}) and system configuration (\autoref{subsec_perf_profiler}) are generated to enhance inference efficiency.

\textbf{Offline.}
To ensure CoServe runs efficiently on various devices, offline profiling is performed once for each device using a set of microbenchmarks to determine the optimal configuration. 
This process establishes the optimal memory allocation and the number of executors for both the CPU and GPU.
Additionally, it evaluates the performance matrix (e.g., latency, memory footprint) to guide online operations and accurately estimates experts' usage probabilities for better system initialization.

\textbf{System initialization.}
After obtaining the configuration information in the offline phase, the executor creator creates the inference executors (Steps 1 to 2 in \autoref{fig4_system_structure}).
Then, the expert initializer within the executor loads the experts into the model pool (Step 3).
Experts are distributed into each executor in a round-robin manner, prioritized by descending usage probabilities, until the memory is fully utilized.


\textbf{Online.}
When a request arrives, it is enqueued in an inference executor’s request queue, awaiting processing.
To minimize the frequency of expert switching, the inference request scheduler utilizes a dependency-aware scheduling method to efficiently assign requests to the appropriate executor and determine their execution order (Steps 4 to 5 in \autoref{fig4_system_structure}).
During execution, the batch splitter dynamically divides the batch of requests based on expert performance and the available memory at that moment (Step 6).

If the required expert is available in the model pool, the inference is executed directly (Step 7).
Otherwise, an expert switching is needed, where the current expert is unloaded from the model pool to free up memory for loading the required expert (Step 8).
To minimize the likelihood of future expert switching, the expert manager utilizes a dependency-aware approach to unload the experts with the lowest probability of future use.

\subsection{Dependency-aware Request Scheduling} \label{subsec_request_scheduling}

The scheduling process is as follows.
First, the request scheduler predicts the additional inference latency for each executor's request queue upon adding a new request.
Then, the request is assigned to the most appropriate executor queue. 
Next, the scheduler arranges the order of the requests in the queue.
Finally, the batch splitter divides the requests into multiple batches during inference for processing.

\begin{figure}[t]
\centering
\includegraphics[width=1\linewidth]{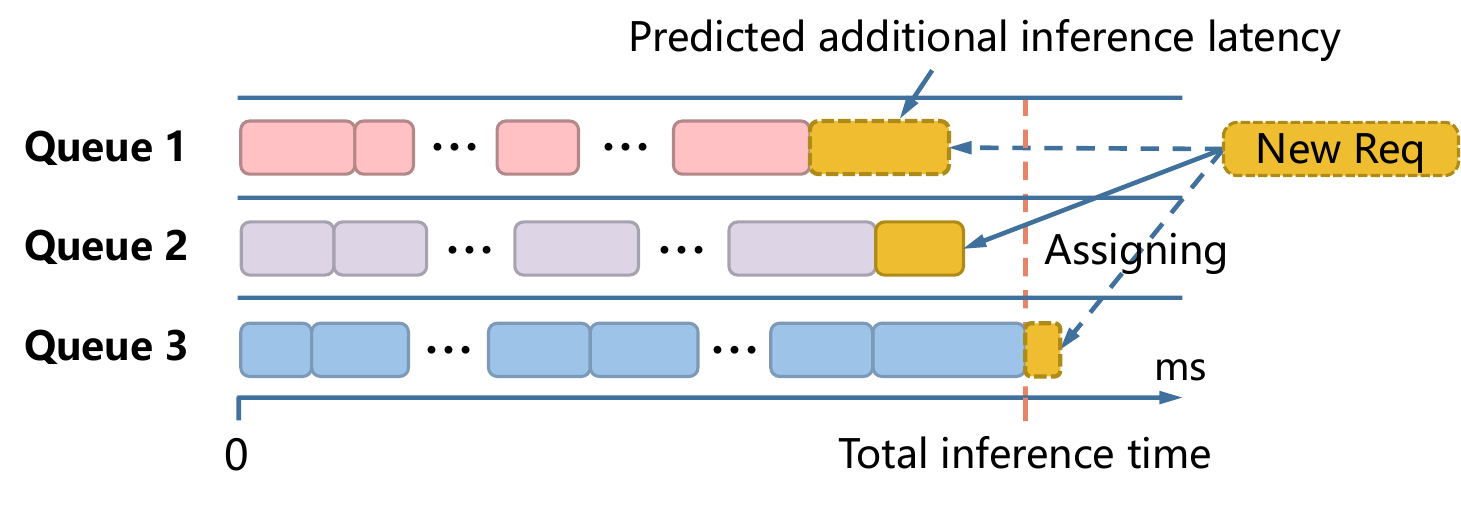}
\caption{Example of request assignment.
The yellow bars represent the predicted additional inference latency after a new request is added to each queue.
The request is assigned to Queue 2, which offers the shortest additional inference latency while minimizing the total inference time.}
\label{fig7_request_dispatch}
\end{figure}

\textbf{Prediction of additional inference latency.}
The additional inference latency consists of execution latency and expert switching latency.

In CoServe, execution latency is estimated as a constant.
This estimation assumes that requests within a batch are processed using the same expert, as CoServe attempts to batch requests utilizing the same expert together.
Specifically, we observe that the overall batch latency scales linearly with the number of requests, expressed as: $latency = K \times (number\ of\ requests\ in\ the\ batch) + B$, provided that all requests in the batch rely on the same expert for processing.
The latency of the first request is $K + B$, while subsequent requests incur a latency of $K$.
Both constants, $K$ and $B$, are precisely measured during the offline phase (details in \autoref{subsec_perf_profiler}).

The expert switching latency is either zero or the time required to load the expert.
It is zero under two conditions.
The first condition occurs when the expert is already present in the model pool, eliminating the need for loading.
The second condition arises when the queue already contains requests utilizing the same expert, allowing the expert to be loaded during the processing of a preceding request.

\begin{figure}[t]
\centering
\includegraphics[width=1\linewidth]{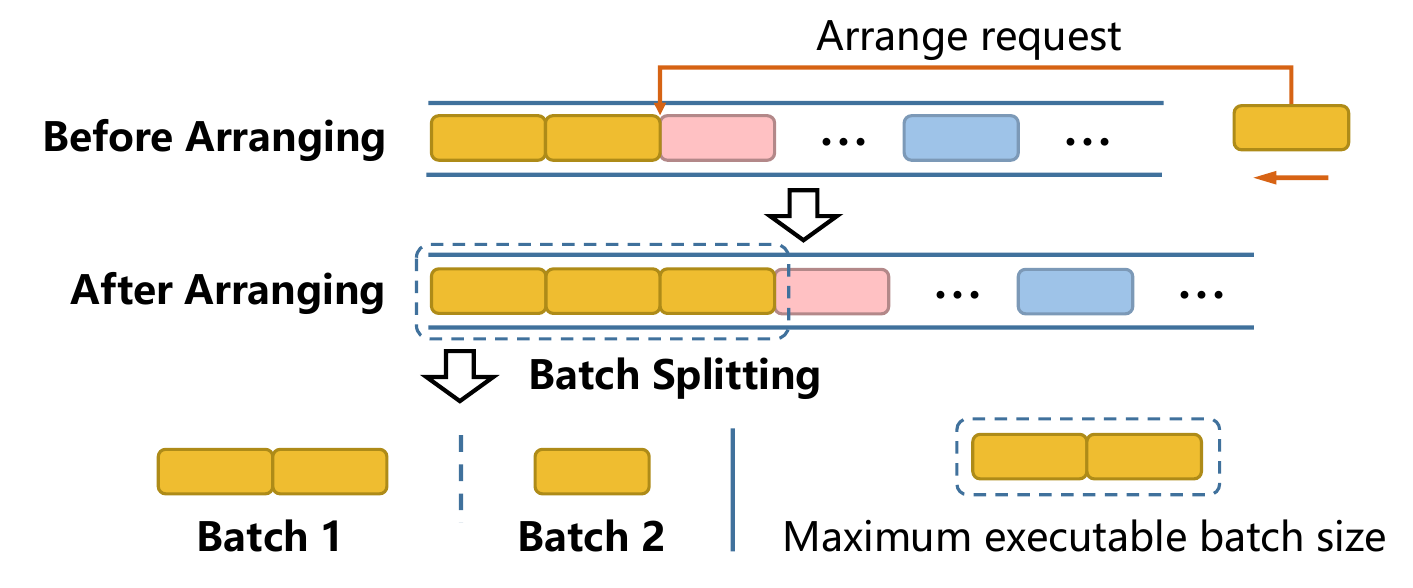}
\caption{Example of request arranging and splitting.
Identical colors represent requests utilizing the same expert.
First, incoming requests are arranged to follow existing requests utilizing the same expert, grouping them together.
Then, these requests are divided into multiple batches based on the current maximum executable batch size for inference.}
\label{fig8_request_scheduling}
\end{figure}

\textbf{Request assigning.}
A task comprises many continuously incoming requests.
To complete the task as quickly as possible, the primary principle for assigning requests is minimizing the current total inference time across all executor queues.
Since executors operate in parallel, the total inference time is determined by the queue with the longest inference time.
For instance, as illustrated in \autoref{fig7_request_dispatch}, the lengths of the queues correspond to their respective total inference times, with Queue 3 dictating the total time.
The yellow bars indicate the additional inference latency incurred when a new request is added to each queue.
Consequently, assigning the request to either Queue 1 or Queue 2 results in minimal total inference time.

When multiple assigning schemes achieve the same minimal total inference time, we select the queue that results in the smallest increase in inference latency for the new request.
In \autoref{fig7_request_dispatch}, the request is assigned to Queue 2.

In summary, the assignment approach minimizes the current total inference time across all executors.
It also preserves more assignment capacity for future requests, enabling more flexible scheduling options.

\textbf{Request arranging.}
Once a request is assigned to a queue, the request scheduler arranges it behind other requests that use the same expert, if such requests are present in the queue.
This groups requests that use the same expert together.
For example, as illustrated in \autoref{fig8_request_scheduling}, requests utilizing the same expert are represented by identical colors.
The incoming requests are arranged to follow existing requests that use the same expert.
This strategy ensures that all requests using the same expert are processed together.
By handling these requests as a group, the expert is loaded at most once, effectively preventing multiple expert switches.

\textbf{Request splitting.}
\liao{The batch size for inference must not exceed the current maximum executable batch size.
The batch splitter is used to enforce this constraint by dividing a set of requests into multiple batches, as shown in \autoref{fig8_request_scheduling}.
}
The current maximum executable batch size is determined by two factors.
The first factor is the largest batch size that the available memory can accommodate.
The second factor is the maximum batch size measured by the performance profiler (see \autoref{subsec_perf_profiler}).
The smaller of these two values is adopted as the current maximum executable batch size.
This batching strategy maximizes the expert’s inference efficiency while considering available resources.

\subsection{Dependency-aware Expert Management} \label{subsec_expert_management}

When the required expert is not available in the model pool, it must be loaded for inference.
If there is insufficient memory to accommodate the new expert, existing experts must be evicted to free up space.
The expert manager utilizes a two-stage eviction strategy that prioritizes the removal of experts with a low likelihood of future use.

\begin{figure}[t]
\centering
\includegraphics[width=1\linewidth]{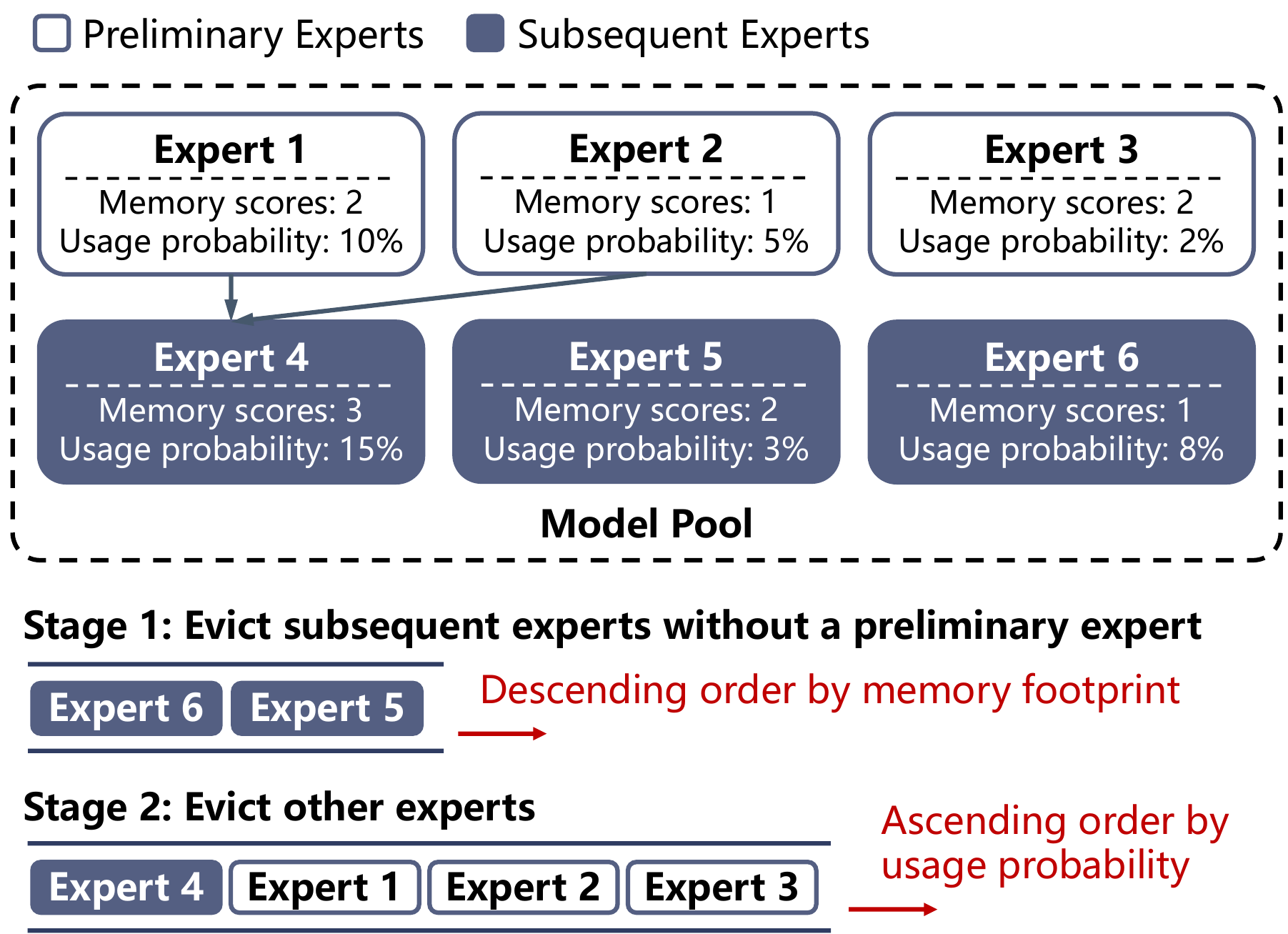}
\caption{Example of the two-stage expert eviction strategy.
The memory scores represent the normalized memory footprint of each expert.}
\label{fig9_model_unloading}
\end{figure}

First, the expert manager prioritizes evicting subsequent experts that lack preliminary dependencies.
Since these experts are not executed until their preliminary experts are fully loaded, they can cause unnecessary memory waste.
As illustrated in Stage 1 of \autoref{fig9_model_unloading}, these experts are sorted in descending order of memory footprint and evicted sequentially until enough memory is available to load the new expert.
This strategy minimizes the number of experts evicted while satisfying memory constraints.

If evicting all such experts does not free sufficient memory, the expert manager evicts experts based on their usage probability.
Experts' usage probabilities can be determined during the offline phase.
As depicted in Stage 2 of \autoref{fig9_model_unloading}, experts are sorted in ascending order of usage probability and then evicted sequentially until adequate memory is available.
This approach ensures that the model pool retains experts with the highest usage probabilities, thereby reducing the likelihood of future expert switching.

\subsection{Efficient Memory Management} \label{subsec_memory_allocation}

Balancing memory allocation between expert loading and inference intermediate results is critically important.
To address the memory trade-off challenges outlined in \autoref{subsec_motiv_memory} for various processors and CoE models, we adopt two adaptive memory allocation strategies tailored to the computational capabilities of each device.
On processors with limited computational performance, we ensure that the memory allocated to inference satisfies the requirements of the maximum batch size.
Conversely, on high-performance processors, inference using the maximum batch size may consume all available memory.
Therefore, it is essential to search for an appropriate allocation that balances the memory footprint between expert loading and intermediate results.

\textbf{Memory allocation under limited computation performance.}
\liao{In processors with limited computational performance, the maximum batch size for experts is usually small and occupies minimal memory.
For such processors, performing inference at the maximum batch size ensures optimal utilization of computational resources, with the remaining memory fully reserved for loading experts.
The maximum batch size is determined in the offline phase (\autoref{subsec_perf_profiler}).
}

\textbf{Memory allocation under sufficient computation performance.}
When the maximum batch size for experts can occupy a substantial portion of the memory, we propose a search strategy that identifies suitable memory allocation.
\liao{The search strategy relies on a CDF (cumulative distribution function) of expert usage, which is generated by the expert usage probabilities obtained offline (details in \autoref{subsec_perf_profiler}).}

\begin{figure}[t]
\centering
\includegraphics[width=0.8\linewidth]{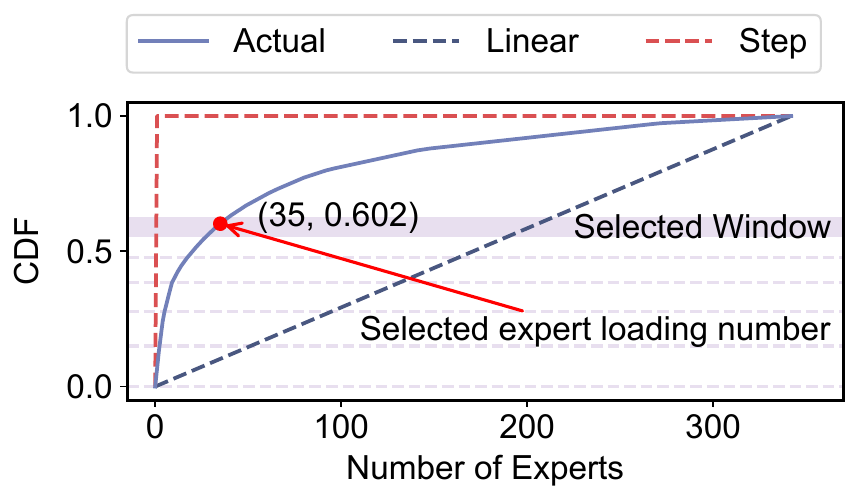}
\caption{Example of cumulative distribution functions (CDF) for expert usage.}
\label{figx_cdf_method}
\end{figure}

We first describe the characteristics of the CDF for expert usage.
There are two extreme scenarios.
The first scenario occurs when all experts have identical usage probabilities.
The second scenario happens when the first expert has a 100\% usage probability, and all other experts have 0\%.
\autoref{figx_cdf_method} illustrates these two cases with linear and step function CDFs.
In real-world situations, experts have varying usage probabilities.
By sorting the experts in descending order of usage probability, the resulting CDF curve falls between the linear and step functions (the Actual curve in \autoref{figx_cdf_method}).

Next, we utilize a decay window approach to identify a suitable amount of memory for experts. 
The core idea is to apply a sliding decay window on the CDF, and then perform sample inference requests at the upper bounds of the window {\color{black} using a smaller, representative dataset sampled from the application scenario.}
The window where the throughput starts to drop is selected and the optimal number of experts is determined within the window.
The dashed horizontal line in \autoref{figx_cdf_method} illustrates the window sliding process and the final selected window.

Initially, the lower bound of the window is set to 0, and the upper bound is the initial window size. 
The decay factor is defined in \autoref{equ_decay}.
\liao{Every time the window slides, its size is reduced by multiplying the original size with the decay factor.
Starting from the first window, CoServe loads experts whose number equals the upper bound of the window and performs sample inference requests by a smaller dataset to generate a throughput value.
Intuitively, the throughput will increase at the start due to more efficient use of computation, but it will drop when the memory contention between intermediate results and experts kicks in.
Therefore, we apply the linear fitting method to predict the upward trend using the first \textit{N} throughput values, as shown in \autoref{equ_linear}.
The window stops sliding when the actual upward trend deviates from expectations (e.g., the throughput starts to decline), as formulated in \autoref{equ_error}.
}

\begin{equation} \label{equ_decay}
\small
    decay\ factor = 1 - \frac{initial\ window\ value}{100}
\end{equation}

\begin{equation} \label{equ_linear}
\small
    f(N) = kN + b
\end{equation}

\begin{equation} \label{equ_error}
\small
    \frac{f(N + 1) - actual\ result}{f(N + 1)} > error\ margin
\end{equation}

\liao{When the sliding process terminates, CoServe randomly selects a value within the window as the optimal number of experts to load, since the decay window gradually narrows the selection space, and differences between values within the window become negligible. 
Once the optimal number of experts is determined, memory is reserved accordingly, with the remaining memory allocated for batch inference. }

\begin{figure}[t]
\centering
\includegraphics[width=1\linewidth]{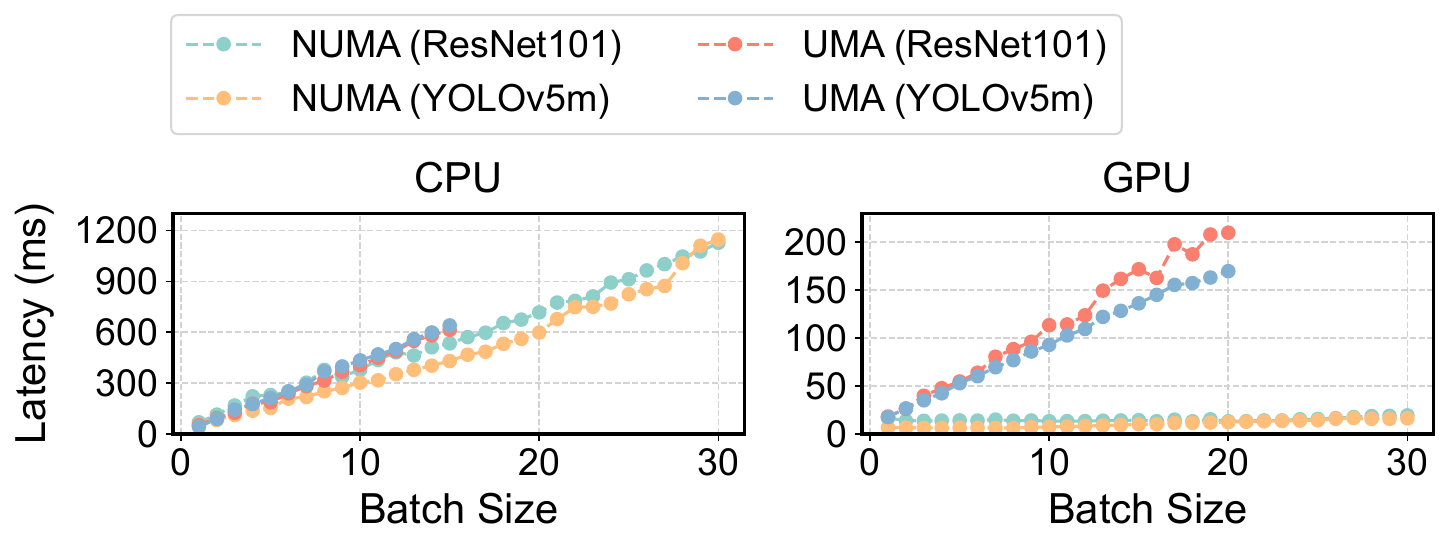}
\caption{Variation of execution latency with increasing batch sizes.}
\label{fig6_multi_batch_latency}
\end{figure}

\subsection{Configuration Information}\label{subsec_perf_profiler}

In the offline phase, CoServe generates configuration information to guide online operations.
Here, we present the configuration information mentioned in previous sections and explain how it is obtained.
The configuration information consists of three components: expert performance metrics, expert information and user-configurable parameters.


\textbf{Expert performance metrics} include maximum batch size, execution latency, and memory footprint, profiled by running the microbenchmarks.
{\color{black} The microbenchmarks leverage real-world samples to reflect the true performance of experts.}
\liao{Experts on GPU and CPU have distinct performance matrices and their performance matrices should be profiled individually.
\textit{It is important to note that experts of the same model architecture are profiled only once}, as their computation complexity (i.e., the number of parameters and floating point operations) is the same.
}

\liao{The \textbf{maximum batch size} is determined by running a microbenchmark with varying batch sizes, and a sample result is illustrated in \autoref{fig3_batch_latency_example}.
It is achieved when the average latency plateaus, indicating that the processor is nearly fully utilized.
It is used for request splitting (\autoref{subsec_request_scheduling}) and memory allocation under limited computation performance (\autoref{subsec_memory_allocation}).
}

\liao{The \textbf{execution latency} is profiled by running the same microbenchmark used to calculate the maximum batch size.
A sample result is present in \autoref{fig6_multi_batch_latency}.
CoServe requires the gradient \textit{K} and intercept \textit{B} on the Y-axis.
This metric is used to assist the prediction of additional inference latency (\autoref{subsec_request_scheduling}).}

\liao{During the profiling of the maximum batch size and execution latency, the \textbf{loading latency} and \textbf{memory footprint} of experts are also recorded.
The loading latency is used to predict the expert switching latency (\autoref{subsec_request_scheduling}).
The memory footprint is normalized to the memory score and used for expert management (\autoref{subsec_expert_management}).}

\textbf{Expert information} comprises routing rules and expert usage probabilities.
\textbf{Routing rules}, provided by the user, are part of the CoE model and determine which experts to handle a given request.

There are two ways to obtain \textbf{expert usage probabilities}.
First, if the routing rules are ambiguous (e.g., they rely on a trained routing model), we can run the CoE routing on a small, real-world sample dataset to record each expert’s usage probability.
Second, if the routing rules are predefined, expert usage probabilities can be calculated directly.
For example, in circuit board inspection, users can specify which components are inspected by which experts.
Because the distribution of component quantities is known, these probabilities are straightforward to compute.
The expert usage probability is used to determine which experts are loaded during initialization (\autoref{subsec_structure}), the order in which experts are evicted (\autoref{subsec_expert_management}), and memory allocation (\autoref{subsec_memory_allocation}).

\textbf{User-configurable parameters} include allocated memory scores and the number of executors.
Although the optimal \textbf{memory allocation} and \textbf{the number of executors} can be determined by running microbenchmarks, users can still manually allocate memory using memory scores and specify the number of executors.

\section{Evaluation} \label{sec_evaluation}

We evaluate CoServe against state-of-the-art CoE serving systems in a real-world intelligent manufacturing scenario. 
We begin with a description of the experimental setups, followed by a detailed presentation of the application results. 
Finally, we conduct ablation studies to analyze the impact of CoServe’s design choices on performance.

\subsection{Evaluation Setting}

\textbf{Hardware.}
We conducted experiments on devices with non-unified memory architecture (NUMA) and unified memory architecture (UMA).
The detailed hardware specifications are provided in \autoref{tab2_hardware}.
CoServe is implemented using PyTorch, and our methods and implementations are designed to be hardware-agnostic.

\textbf{Application.}
Since no publicly available CoE model exists for testing, we built a CoE model in a real-world intelligent manufacturing scenario to conduct circuit board quality inspection. We validate our system in this application.

\textbf{Evaluation metrics.}
{\color{black} In circuit board inspection applications, circuit boards are continuously fed into the inspection system at fixed time intervals, requiring all target component images to be fully analyzed within a specified time frame (e.g., 3 minutes).
However, real-time processing for individual images is not strictly mandatory.
As a result, overall throughput serves as the primary performance metric.}

\begin{table}[t]
\centering
\small
\renewcommand{\arraystretch}{1.1}
\caption{Hardware for evaluation.}
\begin{tabular}{ccc}
\toprule
           & NUMA              & UMA                   \\
\midrule
GPU        & NVIDIA RTX3080Ti     & Apple M2              \\
CPU        & Intel Xeon Silver 4214R & Apple M2              \\
GPU Memory & 12~GB                 & \multirow{2}{*}{24~GB} \\
CPU Memory & 16~GB   & \\
SSD & MTFD-DAK480TDS & APPLE AP0512Z \\
\toprule
\end{tabular}
\label{tab2_hardware}
\end{table}

\textbf{Workload.}
Circuit Board A comprises 352 component types, while Circuit Board B includes 342 component types.
In real-world production, a component image is input every 4 ms.
We delineated the evaluation into four tasks:
\begin{itemize}
    \item \textbf{Task A1:} Continuously input 2,500 requests belonging to Circuit Board A.
    \item \textbf{Task A2:} Continuously input 3,500 requests belonging to Circuit Board A.
    \item \textbf{Task B1:} Continuously input 2,500 requests belonging to Circuit Board B.
    \item \textbf{Task B2:} Continuously input 3,500 requests belonging to Circuit Board B.
\end{itemize}

\textbf{Models.}
A dedicated classification expert is specifically trained for each component to detect defects such as physical damage. 
For some components, after the classification expert confirms no issues, an additional object detection expert is used to further verify alignment points and determine if the soldering direction is correct.
The classification experts are based on the ResNet101~\cite{he2016deep} architecture, with each expert having unique and specialized weights. The object detection experts utilize two architectures: YOLOv5m and YOLOv5l~\cite{YOLOv5_2020}.
Multiple types of components may share the same object detection expert.


\textbf{Baselines.}
To the best of our knowledge, Samba-CoE~\cite{prabhakar2024sambanova} is currently the only system that has explored the deployment of large-scale CoE models.
Therefore, it serves as the most suitable baseline for our evaluation.
Building upon Samba-CoE, we establish three baseline systems.

\begin{enumerate}
    \item \textbf{Samba-CoE:}
    When deploying this baseline on NUMA devices, experts are loaded into GPU memory for inference, utilizing CPU memory as a cache.
    Due to the large number of experts, it is not feasible to load all of them into GPU and CPU memory simultaneously.
    Therefore, during expert swapping, if an expert is present in CPU memory, it is loaded from there; otherwise, it is retrieved from SSD.
    On UMA devices, with a shared CPU-GPU memory architecture, tiered caching is not used, and experts are loaded directly from SSD.
    The expert replacement strategy is LRU, and since Samba-CoE lacks request scheduling optimization, we employ the first-come, first-served (FCFS) approach to handle requests.
    
    \item \textbf{Samba-CoE FIFO:} 
    In this baseline, we modified the expert replacement strategy in Samba-CoE to use the commonly used First-In-First-Out (FIFO)~\cite{deng2020wireless} approach to evaluate its performance.
    
    \item \textbf{Samba-CoE Parallel:} 
    In this baseline, we create multiple parallel inference executors specifically for Samba-CoE to evaluate its performance.
    To ensure a fair and consistent comparison, the number of executors is matched to that of CoServe.
    The incoming inference requests are distributed among different inference executors in a round-robin manner.
\end{enumerate}

\begin{figure}[t]
\centering
\includegraphics[width=1\linewidth]{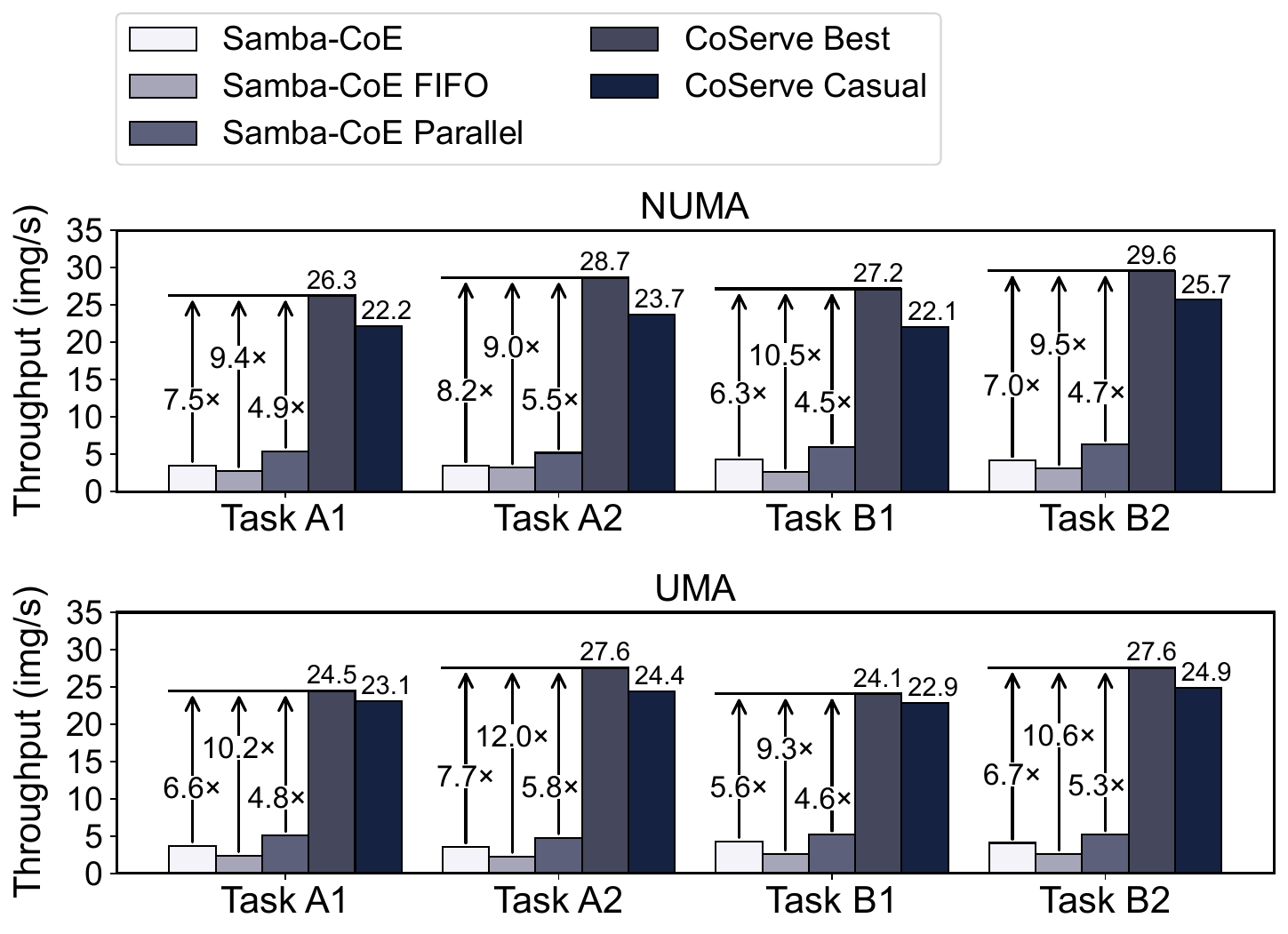}
\caption{Throughput of CoServe and baselines.}
\label{figx_exp1_throughput_comp}
\end{figure}

\subsection{Application Results}

\textbf{Throughput.}
\autoref{figx_exp1_throughput_comp} shows the throughput of CoServe compared to baseline methods across various tasks.
CoServe Best refers to the results obtained by implementing optimal memory allocation strategies and configuring an appropriate number of inference executors.
Overall, compared to Samba-CoE, CoServe achieves a throughput improvement of 4.5$\times$ to 10.5$\times$ over the baselines on NUMA devices and 4.6$\times$ to 12$\times$ on UMA devices.

CoServe Casual refers to a casually selected memory allocation and the number of executors.
The specific configuration for CoServe Casual is as follows.
On the GPU of NUMA devices, 75\% of the memory is allocated for expert loading and 25\% for batch inference.
We configured three GPU executors for NUMA devices due to their robust computational capabilities and two GPU executors for UMA devices.
Additionally, both NUMA and UMA devices are equipped with one CPU executor each.

Compared to the baselines, CoServe Casual achieves up to 8.5$\times$ and 10.6$\times$ improvements in throughput on NUMA and UMA devices, respectively.
This indicates that even if users do not meticulously search for the optimal configuration and instead use intuitive settings, CoServe can still deliver excellent performance.

Compared to CoServe Best, CoServe Casual shows a minimum throughput decrease of 5.71\% on NUMA devices, dropping from 24.5 to 23.1, and a maximum decrease of 18.75\% on NUMA devices, from 27.2 to 22.1.
These results demonstrate that selecting optimal memory allocation and the number of executors during the offline phase can effectively improve inference performance.

\begin{figure}[t]
\centering
\includegraphics[width=1\linewidth]{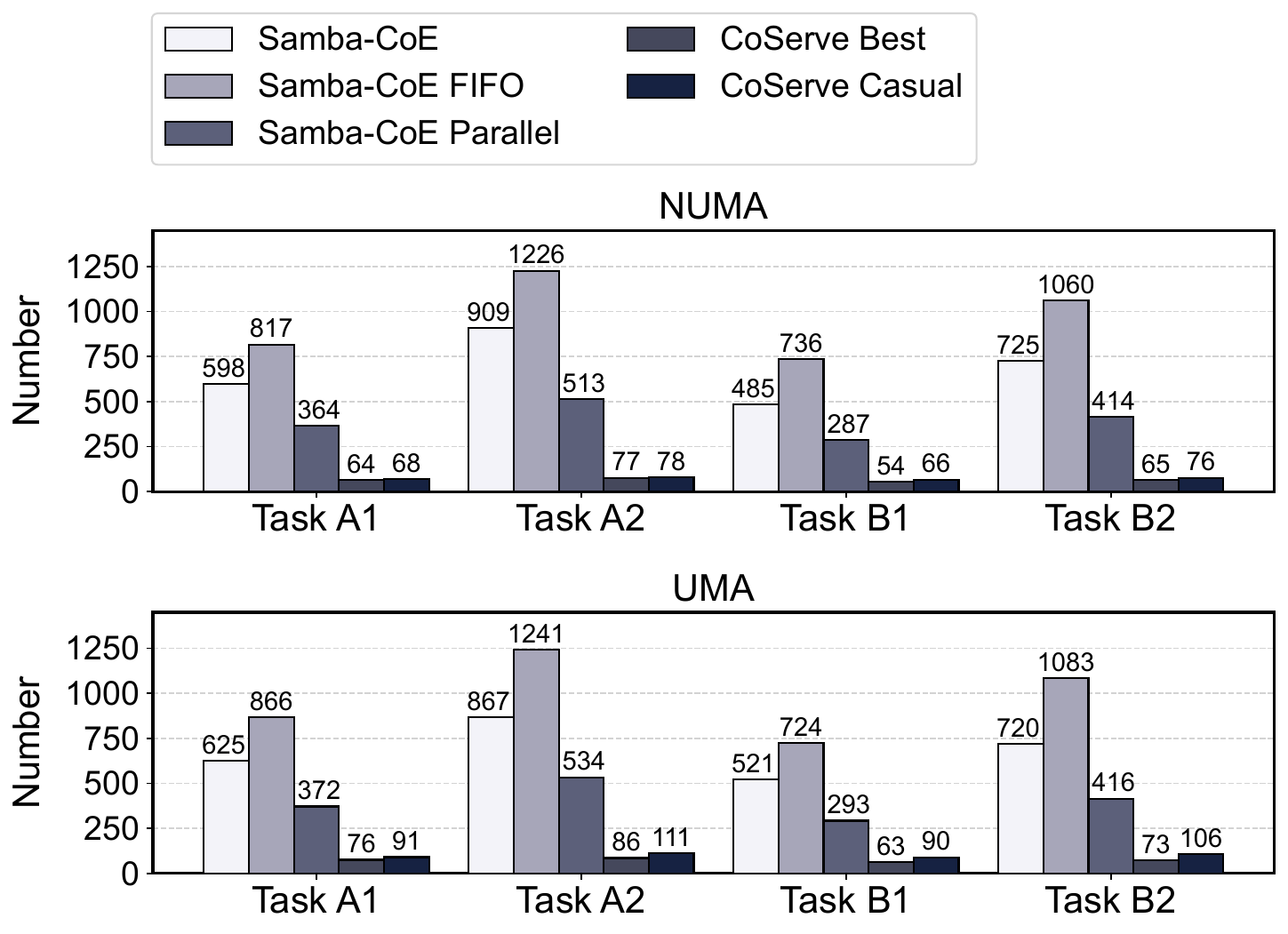}
\caption{Number of expert switches for CoServe and baselines.}
\label{figx_exp2_model_loading_time_compare}
\end{figure}

\textbf{Expert switching.}
\autoref{figx_exp2_model_loading_time_compare} compares expert switches between CoServe and baseline methods.
CoServe reduced expert switching by up to 93.87\%, from 1,060 to 65 in Task B2 on NUMA devices, and by at least 78.5\%, from 293 to 63 in Task B1 on UMA devices.
These results demonstrate CoServe's significant improvement in inference efficiency by minimizing expert switches.

Compared to CoServe Casual, CoServe Best significantly reduces the number of expert switches.
On NUMA devices, the reduction ranges from 1.28\% to 18.18\%, while on UMA devices, it ranges from 16.48\% to 31.13\%.
This indicates that selecting optimal memory allocation and the appropriate number of executors during the offline phase can further reduce the number of expert switches.

\subsection{Ablation Studies}

\begin{figure}[t]
\centering
\includegraphics[width=1\linewidth]{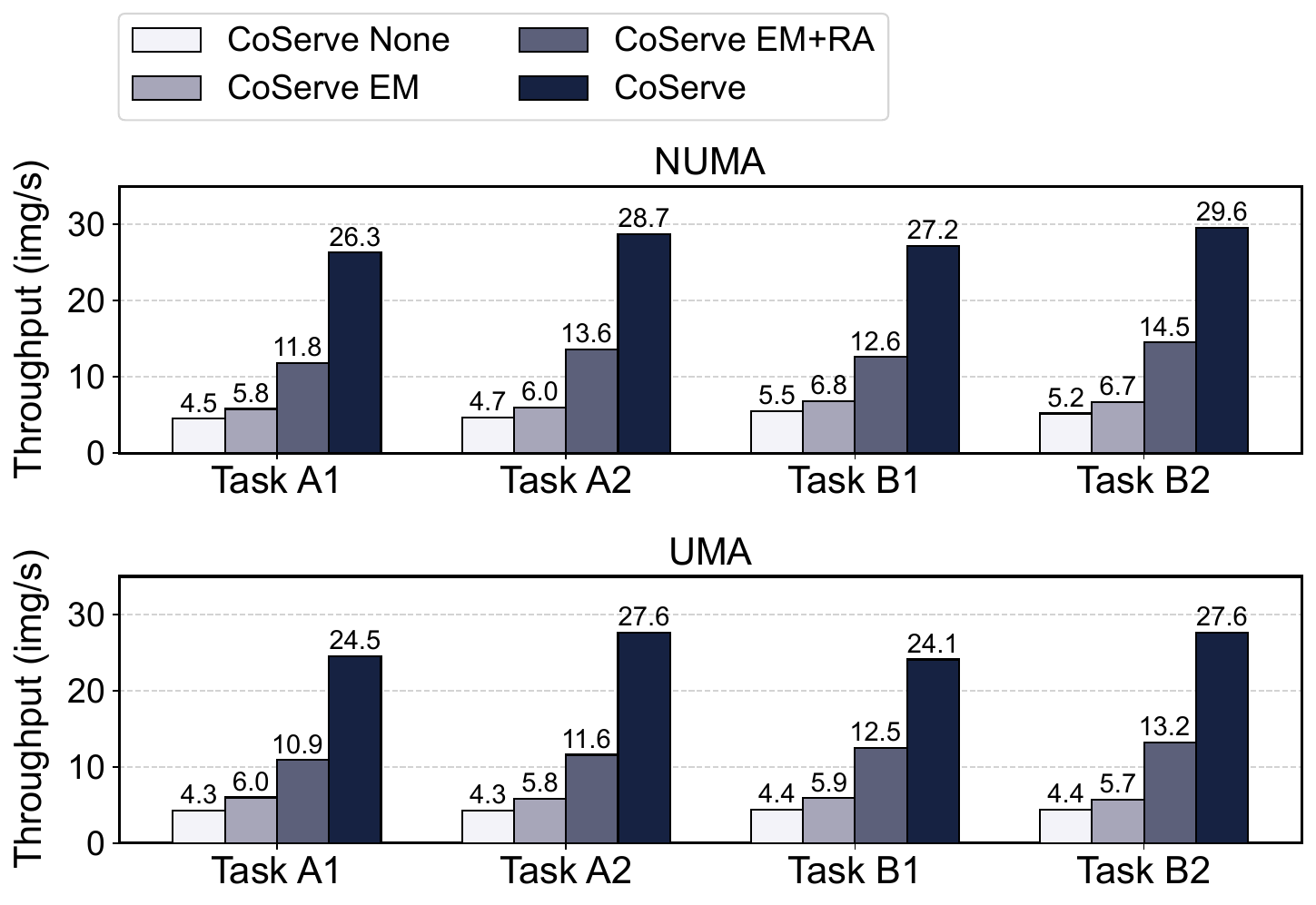}
\caption{Breakdown of throughput for each optimization in CoServe.}
\label{figx_exp5_ablation_throughput_comp}
\end{figure}

\textbf{Performance breakdown.}
\autoref{figx_exp5_ablation_throughput_comp} provides a detailed breakdown of the throughput improvements achieved by expert management and request scheduling.
Request scheduling is further divided into request assigning and request arranging.
CoServe None denotes the baseline system with no optimizations, which utilizes a FIFO strategy for expert replacement and request execution, distributing requests evenly across executors.
CoServe EM introduces expert management optimization, while CoServe EM+RA further incorporates request arranging.
The fully optimized system, labelled CoServe, integrates all optimizations: expert management, request arranging, and request assigning.
The results indicate that each of these optimizations contributes to a significant increase in throughput.

\begin{figure}[t]
\centering
\includegraphics[width=1\linewidth]{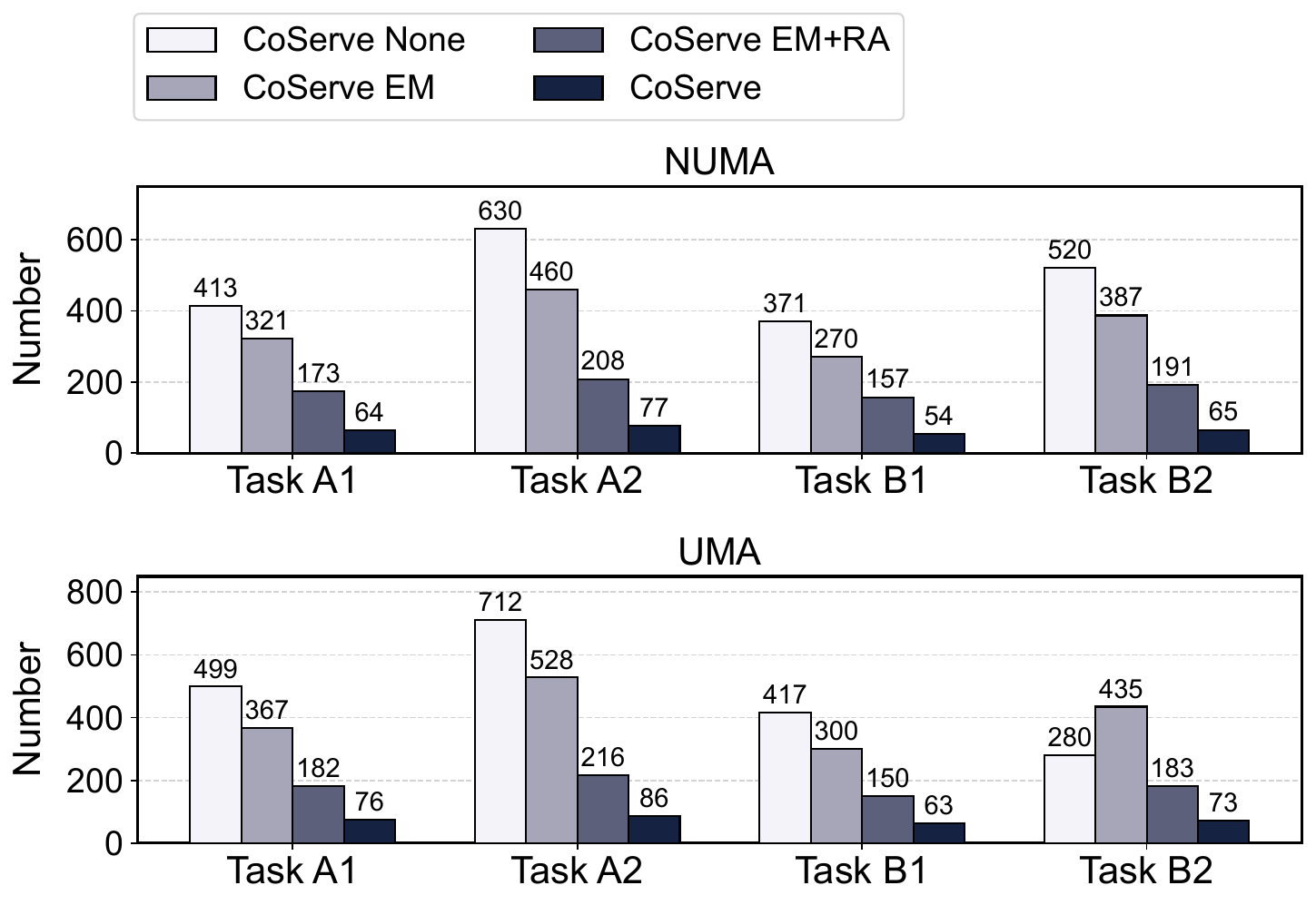}
\caption{Breakdown of the number of expert switches for each optimization in CoServe.}
\label{figx_exp6_ablation_model_loading_number_compare}
\end{figure}

The number of expert switches is shown in \autoref{figx_exp6_ablation_model_loading_number_compare}.
Each optimization reduces the number of expert switches, with the reduction proportionally related to the increase in throughput.
This indicates that CoServe's optimizations effectively reduce the frequency of expert switching, thereby improving throughput.

\begin{figure}[t]
\centering
\includegraphics[width=1\linewidth]{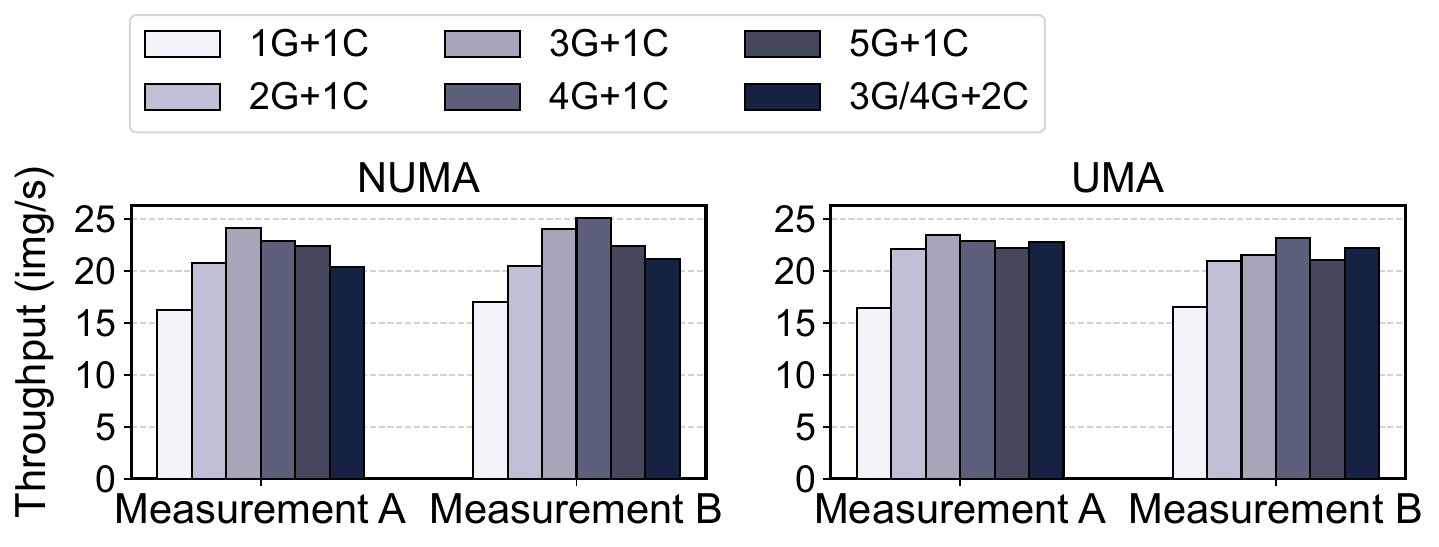}
\caption{Throughput under different numbers of executors.
G and C represent GPU and CPU executors.}
\label{fig13_exp4_worker_number_compare}
\end{figure}

\textbf{Performance with different numbers of executors.}
In the offline phase, we use a portion of the data to test system throughput under different numbers of executors to select the optimal configuration.
The results are shown in \autoref{fig13_exp4_worker_number_compare}.
In the figure, G and C represent the number of GPU and CPU inference executors, respectively.
As the number of CPU inference executors increases, the number of GPU inference executors is kept at the configuration that previously achieved the highest throughput.
Some of these configurations include 3 GPU executors, while others have 4.
On both NUMA and UMA devices, the configuration of 3 GPU executors and 1 CPU executor yields better performance for Task A, while 4 GPU executors and 1 CPU executor perform better for Task B.
When the number of executors decreases, insufficient utilization of computational resources leads to performance degradation.
Conversely, when the number of executors increases, the additional overhead introduced also results in performance loss.

\begin{figure}[t]
\centering
\includegraphics[width=1\linewidth]{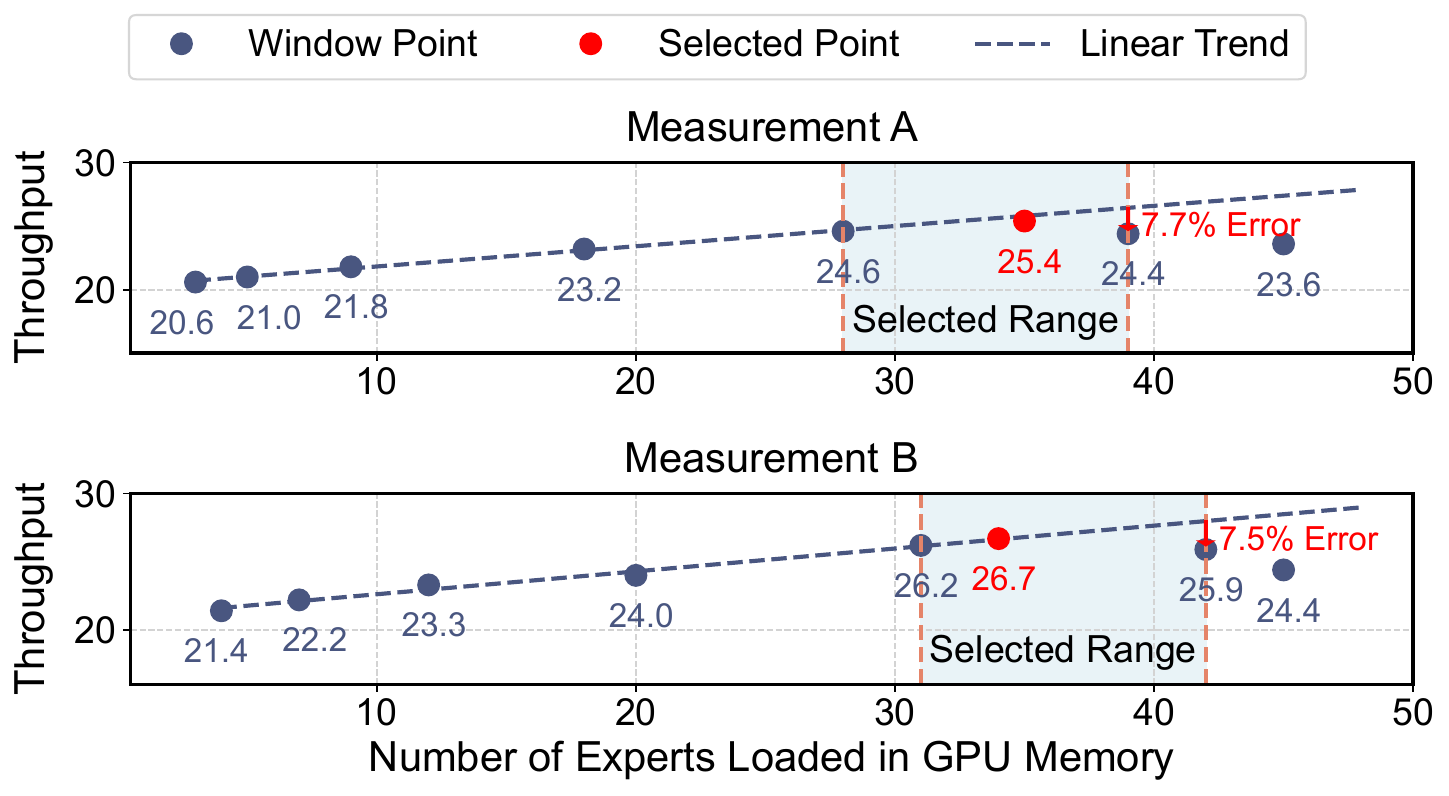}
\caption{Throughput measured at the window boundaries during the sliding window process.}
\label{fig13_exp4_loaded_number_compare}
\end{figure}

\textbf{Performance of different memory allocation.}
We applied the method described in \autoref{subsec_memory_allocation} to find an appropriate selection window on the GPU of the NUMA device.
The initial window size was set to 15, with a linear error rate of 5\%.
\autoref{fig13_exp4_loaded_number_compare} shows the throughput achieved as the number of loaded experts changes, corresponding to the upper and lower bounds of the sliding and decaying window.
As the number of loaded experts increases, throughput first rises and then falls.
For Task A, the final selection window corresponds to loading 28 to 39 experts, with a linear error of 7.7\%.
We chose to load 35 experts, achieving a throughput of 25.4.
For Task B, the final selection window corresponds to loading 31 to 42 experts, with a linear error of 7.5\%.
We chose to load 34 experts, achieving a throughput of 26.7.
The fact that the peak throughput lies within the selected window demonstrates the effectiveness of our approach.

\begin{figure}[t]
\centering
\includegraphics[width=1\linewidth]{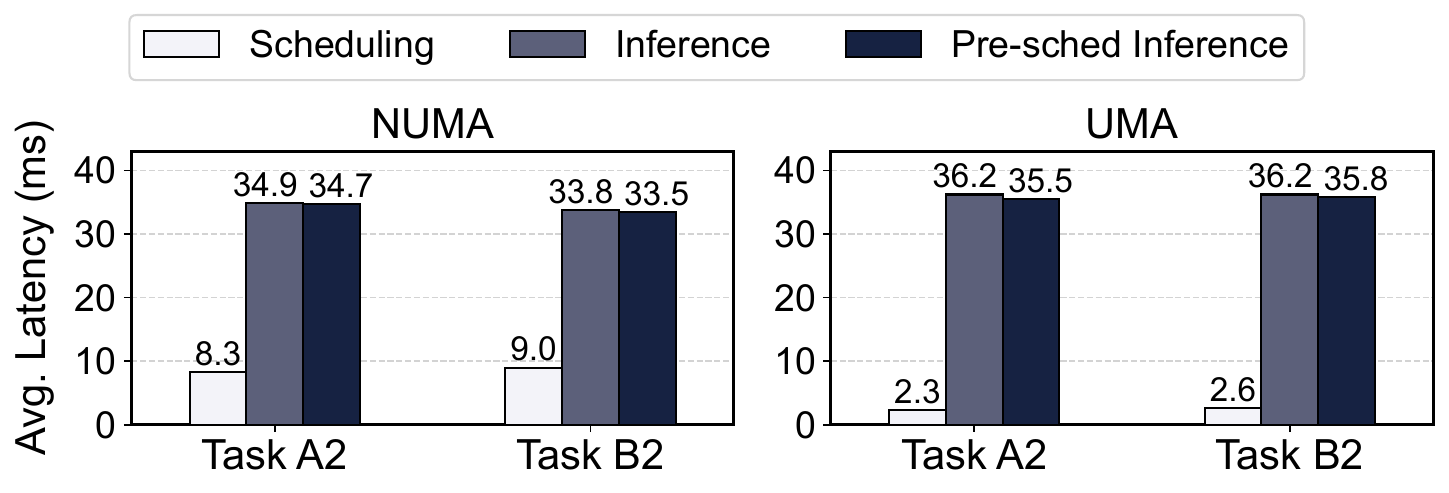}
\caption{Average latency of request scheduling, inference, and pre-sched inference for a single request.}
\label{figx_expx_request_overhead}
\end{figure}

\textbf{Overhead analysis.}
\liao{We study the overhead of CoServe by comparing the average latency of request scheduling with inference latency.
Results are shown in \autoref{figx_expx_request_overhead} and demonstrate that the latency of request scheduling is shorter than inference latency.
In CoServe, request scheduling is performed by the CPU and occurs in parallel with inference processes in GPU executors.
Since the scheduling is faster than inference, it does not cause any bottleneck or negatively impact GPU inference performance.
}

\liao{However, it still remains unclear whether the scheduling will significantly affect the CPU inference and further the overall performance.
We conduct additional experiments by scheduling requests offline in advance and performing inference directly without any scheduling.
This setup, referred to pre-sched inference (\autoref{figx_expx_request_overhead}), achieves the same request sequence as CoServe but with zero scheduling overhead.
The performance difference between regular inference and pre-scheduled inference quantifies the impact of scheduling on overall performance. 
We observe that the performance gap is less than 3\%. 
These findings indicate that request scheduling has minimal impact on inference performance.
}

We measured the proportion of time spent on expert management relative to the total task time for each executor and calculated the average.
The time spent on expert management does not exceed 0.2\%.
However, it leads to an improvement in throughput, making this overhead worthwhile.

\section{Related Work}

\subsection{CoE and MoE Model}

The multi-expert models are categorized into two types: Collaboration of Experts (CoE) and Mixture of Experts (MoE).
Representative CoE models include Samba-CoE~\cite{prabhakar2024sambanova}, Qihoo 360 CoE~\cite{huang2024ccoe} and Bench-CoE~\cite{wang2024bench}, while MoE models are represented by Mixtral-MoE~\cite{jiang2024mixtral}, JetMoE~\cite{shen2024jetmoe}, Jamba~\cite{lieber2024jamba}, DeepSeekMoE~\cite{dai2024deepseekmoe}, and Branch-Train-Mix~\cite{sukhbaatar2024branch}.

The differences between CoE and MoE models are as follows.
CoE models consist of multiple independent models, where each model is an independently trainable expert.
Experts in CoE models can be easily managed individually, and they can be added or removed independently.
MoE models, on the other hand, are composed of multiple experts within a single model, and training must occur as a whole.
The router in CoE models can be manually adjusted, while in MoE models, the router is generated during training and cannot be optimized independently.
Our work focuses on inference optimization for CoE models.


\subsection{Multi-Expert Inference Optimization}

\textbf{CoE model inference optimization.}
Samba-CoE~\cite{prabhakar2024sambanova} optimizes CoE model inference by offloading experts to DDR and loading them into HBM as needed, using an LRU strategy for replacement.
However, there is still considerable room for improvement in request scheduling, expert management, and memory management to enhance system efficiency.
Shi \textit{et al.}~\cite{shi2020multi} use multithreading to deploy experts on both NPU and CPU simultaneously, boosting inference performance. 
However, they do not address the memory limitations that prevent loading a large number of experts at once.

\textbf{MoE model inference optimization.}
For MoE model inference, many studies have optimized routing~\cite{kudugunta2021beyond, huang2023towards, hwang2024pre}, expert scheduling~\cite{yi2023edgemoe, eliseev2023fast, kamahori2024fiddler, kong2024swapmoe}, quantization~\cite{kim2023mixture, frantar2023qmoe, yu20248}, and distributed inference~\cite{li2023accelerating, yao2024exploiting}.
Our work specifically aims to optimize the inference efficiency of CoE models and is orthogonal to these approaches.


\section{Conclusion}

In the inference of CoE models with a large number of experts, limited memory on resource-constrained edge devices leads to frequent expert switching, which significantly degrades inference performance.
To address this critical issue, we propose CoServe, which effectively reduces expert switching through dependency-aware request scheduling and expert management.
Additionally, offline measurements fully exploit device performance.
Compared to state-of-the-art baselines, CoServe significantly improves throughput.

{\color{black}
Due to the limited availability of models and test data, CoServe has been validated only in the circuit board defect detection application within the intelligent manufacturing scenario.
However, for other CoE models, as long as the necessary routing module and expert models required by CoServe are provided, its design remains equally applicable.}
We hope our work offers valuable insights for the application, deployment, and serving of CoE models.


\begin{acks}
This work was supported by the National Key R\&D Program of China under Grant No. 2023YFB4503100, the National Natural Science Foundation of China under Grant No. 62272026 and No. 62302257, and the State Key Laboratory of Complex \& Critical Software Environment under Grant No. CCSE-2024ZX-10.
We sincerely thank our shepherd, Antonia Zhai, for her invaluable support and guidance, as well as the anonymous reviewers for their comments and suggestions, which have greatly enhanced the quality of this paper.
\end{acks}

\bibliographystyle{plain}
\balance
\bibliography{reference}

\end{document}